# Fuzzy Multi-Agent Simulation of COVID-19 Pandemic Spreading

**Didier El Baz[1] and Andrei Doncescu[1,2]**

[1]LAAS-CNRS UPR 8001, Toulouse, France,
[2]French West Indies University, 97157, Pointe-à-Pitre, Guadeloupe,
elbaz@laas.fr, andrei.doncescu@laas.fr

In this paper, we present a new approach for Covid-19 Pandemic spreading simulation based on fuzzy multi agents. The agent parameters consider distribution of the population according to age, and the index of socio-economic fragility. Medical knowledge affirms that the COVID-19 main risk factors are age and obesity. The worst medical situation is caused by the combination of these two risk factors which in almost 99% of cases finish in ICU. The appearance of virus variants is another aspect parameter by our simulation through a simplified modeling of the contagiousness. Using real data from people from West Indies (Guadeloupe, F.W.I.), we modeled the infection rate of the risk population, if neither vaccination nor barrier gestures are respected. The results show that hospital capacities are exceeded, and the number of deaths exceeds 2% of the infected population, which is close to the reality.

## 1. Introduction

COVID-19 is an unprecedented pandemic outbreak by the high rate of infection and the global spreading. Besides comping with possible physical illness, a pandemic outbreak can cause employees anxiety and fear. Neither country was not prepared to face this outbreak, especially for low and middle-income countries the number of ICU beds are insufficient to combat Covid-19 pandemic [Ma & Vervoort 2020].

Mathematical and numerical simulation models could be a helpful decision tool to manage the crisis by predicting the number of infections [Currie *et al.* 2020]. Many of these tools consider analytical-based models, using for example, the SIR epidemic differential equations [Wu *et al* 2020], [ Roda *et al.* 2020],[ Giordano *et al.* 2020],[ Ivorra *et al.* 2020] or linear regression [ Rustam *et al.* 2020]. Using deep learning [Chimmula et Zhang 2020] or fuzzy neural networks [Al-qaness *et al.* 2020] is possible to have a data-driven approach. The effects of the lockdown during the pandemic have been analyzed by estimating the number of deaths [Roux *et al.* 2020], [Flaxman *et al.* 2020].

In this paper we present a Multi Agent-based System (MAS), which incorporates several parameters as risk factors. Elder age is the main risk factor for the severe or critical case of infected [Zhou *et al*. 2020]; [Wu & McGoogan 2020]; [Williamson *et al*. 2020]; and obesity (measured by BMI) seems to be the second main risk factor; [Muscogiuri *et al.* 2020]; [Kass *et al.* 2020]; [Lighter *et al*. 2020]. The cited articles the risk factors are introduced in the form of probabilities. driven by medical and statistical knowledge.

In comparison, in this article, we are interested in the number of critical cases due to the risk factors, by handling the uncertainty using fuzzy logic (Fuzzy Multi-Agent Simulation).

This approach seems to be in a context where many countries do not have enough Intensive Care Units (ICU) to manage the crisis[Ma & Vervoort ,2020].

## 2. Fuzzy Multi-Agent Simulation

We adapted the model, implemented in NetLogo platform, epiDEM Travel and Control [Yang & Wilensky, 2011*b*]. In our model the agents have two main characteristics: age and obesity. To do that the population has been classified in 3 fuzzy groups (mild, severe, critical) according to their age and their body mass index BMI (Fig.1) :

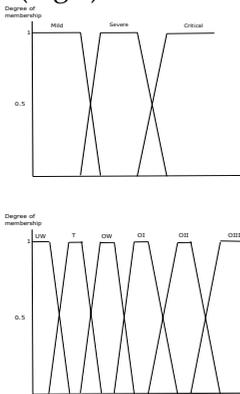

Fig.1 Membership Functions : age and obesity

The elder age is the main risk factor for the severe or critical case and obesity is the second main risk factor. According to demographic data, the population is divided into three age groups: young people, adults, and elderly. The severity of COVID-19 disease roughly corresponds to these three groups [Zhou & et al.,2020], [Williamson *et al.* 2020]. Statistically, the covid-infected young people have a mild form of the disease, adults have a more seriousform and correspond to the group of severe patients, and elderly people represent most critical cases. Obviously, this classification does not consider comorbidities.

We introduced in our model the variants of this virus, delta, which is at least 50% more contagious than the "classic" COVID-19 [Volz *et al.*, 2021], [Davies *et al.*,2021]. Therefore, we defined in our MAS simulation a zone of transmissibility. The transmissibility zone is represented by a bigger centroid around the infected agent than in the case of a non-infected agent.

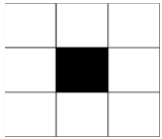

Fig.2 Transmissibility Zone

### 3. Results

The fuzzy MAS simulation presented in this article supposed a missing situation of social distancing or barrier gestures. Therefore, the agents can circulate freely transmitting the disease to everyone with whom it comes into contact.Regarding the characteristics of COVID-19, the parameters are adapted to be directly linked to the disease. Average-recovery-time representing the average duration of the disease has been set to 25 [37 in Zhou & et al., 2020]. Fig.3 displays a comparison of real number of infected people per day (orange curve) and simulated number via our method (blue curve).

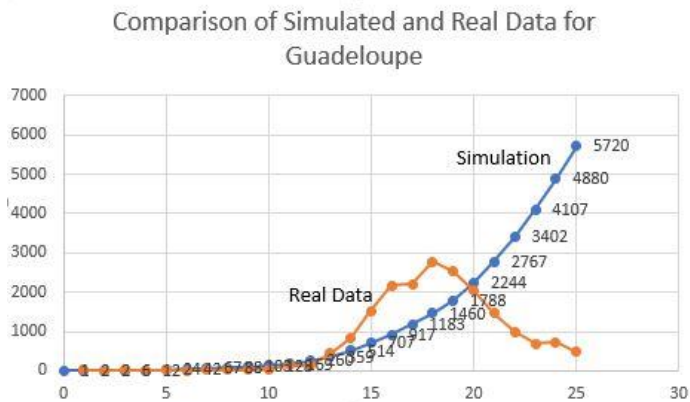

Fig.3 Number of Infected People per Day in Guadeloupe

We know that the lockdown started the day 19 and our simulation shows what it could happened without lockdown.

## 4. Conclusion

In this paper, we presented a simulation model of COVID-19 spreading in an insular context, considering a non-respect of social distancing and barrier gestures. The model of the spread of the COVID-19 pandemic used a MAS and was based on demographic data and medical knowledge based on two risk factors: age and obesity. These two aggravating risk factors were defined as parameters of agents by using fuzzy logic: fuzzy sets and aggregation fuzzy operators.

Further the agents will be defined by considering the vaccination as barrier gesture. In addition, new variant of the virus more contagious will be modelled in this Fuzzy-MAS simulator.

**Acknowledgments:** The authors of this article would like to thank the Agence Régionale de Santé de Guadeloupe (Regional Health Agency of Guadeloupe) and specially Service Analyse des Données de Santé de la Direction d'Evaluation et de Réponse aux Besoins des Populations (Health Data Analysis Department of the Department of Assessment and Response to Populations' Needs) for the provision of epidemiological data (incidence rate).